\documentclass{webofc}
\usepackage[varg]{txfonts}
\usepackage{graphicx}
\usepackage{epsfig}
\usepackage{color}
\usepackage{subfig}
\usepackage{sistyle}
\usepackage{mathtools}
\usepackage{calc}
\usepackage{pgfplots}
\usepackage{tikz}

\usepackage{amssymb,amsmath}

\begin{document}

\woctitle{Slow dynamics of the contact process on complex networks}

\title{Slow dynamics of the contact process on complex networks}

\author{G\'eza \'Odor \inst{1} \fnsep\thanks{\email{odor@mfa.kfki.hu}}}
\institute{Research Centre for Natural Sciences,
Hungarian Academy of Sciences, MTA TTK MFA, 
P. O. Box 49, H-1525 Budapest, Hungary}

\abstract{
The Contact Process has been studied on complex networks exhibiting different
kinds of quenched disorder. Numerical evidence is found for Griffiths phases 
and other rare region effects, in Erd\H os R\'enyi networks, leading rather 
generically to anomalously slow (algebraic, logarithmic,...) relaxation. 
More surprisingly, it turns out that Griffiths phases can also emerge 
in the absence of quenched disorder, as a consequence of sole topological 
heterogeneity in networks with finite topological dimension. 
In case of scale-free networks, exhibiting infinite topological dimension,
slow dynamics can be observed on tree-like structures and a superimposed
weight pattern. In the infinite size limit the correlated subspaces
of vertices seem to cause a smeared phase transition.
These results have a broad spectrum of implications for propagation
phenomena and other dynamical process on networks and are relevant
for the analysis of both models and empirical data.}

\maketitle

\section{Introduction}

Nonequilibrium systems have been a central research topic of statistical 
mechanics \cite{marro1999npt,odorbook,Henkel}. As in equilibrium they
exhibit various phase transitions, critical phenomena and universality 
classes. Since they cannot be treated by thermodynamical formalism
universality provides a guidance's in the zoo of models exhibiting diverging
correlation length and scaling behavior \cite{rmp}.
Power-law scaling behavior has been reported in many real and model systems
for spontaneous occurrence the theory of self-organizing-criticality
was proposed \cite{Bak}. This, however requires the competition of a slow
accumulation and fast dissipation mechanism. Diverging correlation length
on the other hand is generated naturally in nonequilibrium systems by
the currents or fields acting on them. Furthermore, quenched disorder may
also cause extended regions in the parameter space with non-universal 
power-law dynamics \cite{Vojta}.
A fundamental dynamical system model is the Contact Process (CP) 
\cite{harris74,liggett1985ips}, in which sites can be either occupied 
(infected) or empty (susceptible).
By changing the infection rate of the neighbors $\lambda/k$, where $k$ is the
degree of the vertex, a continuous phase transition occurs at the $\lambda_c$
critical point from inactive to active steady state.
The inactive state, characterized by the order parameter, the density of 
infection ($\rho$) is zero. This is also called absorbing state, because 
no spontaneous activation of sites is allowed.

In the new century the interest is shifting from models, defined on Euclidean,
regular lattices to processes defined on general networks 
\cite{barabasi02,mendesbook}.
Since the introduction of a simple model describing the emergence of scaling in 
random networks \cite{Barabasi1999} the study of complex networks is 
flourishing. Multidisciplinary applications involve, for example, the  
WWW, various biological, sociological and technological networks.
These are, mostly scale-free (SF) networks, exhibiting $P(k)\sim k^{-\gamma}$ 
degree distribution of the nodes 
(for recent reviews see \cite{barabasi02},\cite{dorogovtsev07}).
Other families of complex network models are those composed of a $d$-dimensional 
regular lattice and additional long edges \cite{kleinberg}.  
These arise in the context of conductive properties of linear polymers 
with cross-links that connect remote monomers \cite{cc}, in public traffic 
systems \cite{opttrans}, in the case of nanowires \cite{HBU09}, in social
\cite{socimb} and phone networks \cite{phonimb} or to describe forest fires 
\cite{Zekri} among other examples.  
In these models a pair of nodes separated by the distance $l$ of the base lattice
are re-connected by a long edge with the asymptotic probability for large $l$:
\begin{equation}\label{pdis}
p(l) = \beta l^{-s} \ .
\end{equation}
In the special case $s=0$, edges exist with a length-independent probability, 
as in small world networks, therefore these models are called Generalized Small 
World networks (GSW). The $s=2$ case is important from network optimization 
point of view \cite{opttrans} for example.
If $s\ge 2$, they are characterized by a finite topological dimension $d$, 
i.e. $N(l) \sim l^d$, where $N(l)$ is the number of nodes within the graph
distance $l$ from a given node.
Note that if the links vary quickly with respect the time scale of the 
dynamical process defined on these networks they realize L\'evy flights 
\cite{Mo77,hinrichsen}. Usually these annealed networks exhibit different 
behavior than the quenched ones.

Although many network models exhibit infinite topological dimension ($d$), 
simple mean-field approximations cannot capture several important features 
\cite{Castellano2006,Castellano2008,boguna09,GDOM12,LSN12,wbacikk}. 
Very recently it has been conjectured \cite{GPCP,odor172,Juhasz2011fk} that 
generic slow (power-law, or logarithmic) dynamics is observable only in 
networks with finite $d$. This claim is relevant in the light of recent 
developments of dynamical processes on complex networks such as 
the simple model of ``working memory'' \cite{Johnson}, brain dynamics 
\cite{Chialvo}, social networks with heterogeneous communities  \cite{Castello}, 
or to understand the slow relaxation in glassy systems \cite{Amir}.
Slow dynamics has been shown to originate from the bursty behavior of the
agents connected by small world networks resulting in memory effects \cite{KK11}.
On the other hand it can also be related to arbitrarily large ($l<N$), 
correlated rare-regions (RR), which possess long lifetime in the inactive 
phase, above the pure critical point $\lambda_c^0 < \lambda < \lambda_c$.
This can be understood by non-perturbative methods
\cite{PhysRevLett.59.586,sethna88,PhysRevB.54.3328,PhysRevLett.69.534,Monthus,IK11}.

More recently the possibility of power-law dynamics of CP has been investigated
on different BA networks with $\gamma=3$ \cite{BAGP,wbacikk}.
Extensive simulations showed $\rho(t) \propto 1/(t\ln(t))$ density decay and 
$\rho(\lambda,t\to\infty)\propto |\lambda-1|$ steady state behavior
with logarithmic corrections in agreement with the HMF approximations.
On loop-less BA trees the epidemic propagation slows down and a 
nontrivial critical density decay emerges $\rho(t,\lambda_c) \propto t^{-0.5}$.
Additionally, when $k$ dependent weighting was applied, which suppress hubs or 
make the network disassortatative GP-like dynamics was observed in the 
simulations. However, systematic finite scaling study revealed that 
these power-laws saturate in the $N\to\infty$ thermodynamic limit, 
suggesting smeared phase transitions known from Euclidean, regular systems 
if the correlated subspaces can undergo phase transitions themselves, 
when they are effectively above the lower critical dimension of the 
problem: $d_{RR} > d_c^-$ \cite{Vojta}. In this case, the dynamics of the 
locally ordered RR-s completely freezes, and they develop a truly static 
order parameter. Clearly in infinite dimensional networks such RR-s can 
be embedded as a percolation analysis confirmed this \cite{BAGP}.

\section{Optimal fluctuation theory}
\label{sec:OFT}

The basic idea of the optimal fluctuation theory is that the long-time
decay of the order parameter ($\rho(t)$) is dominated by the regions
of size $l$, which are rare in general: $P(l) \propto \exp(-c l)$, but can 
exhibit exponentially long lifetimes $\tau(l) \propto \exp(b l)$.
In particular for the the density of infected sites of CP, RR-s provide 
the leading order contribution in the inactive phase: 
\begin{equation}
\rho(t) \sim \int l P(l) \exp(-t/\tau) dl  \ ,
\end{equation}
which in the saddle point approximation results in $\rho(t)\sim t^{-c/b}$
decay, with a non-universal, $\lambda$ dependent exponent \cite{GPCP}. 
This is the so called Griffiths Phase (GP) \cite{Griffiths,Vojta}, 
bordered by a stretched exponential decay law at the critical point 
$\lambda_c^0$ of the clean system, and by the dirty critical point 
$\lambda_c>\lambda_c^0$, where the evolution becomes logarithmically slow
\begin{equation}
\rho(t) \sim \ln(t/t_0)^{-\tilde\alpha} \ ,
\label{logdec}
\end{equation}
thanks to $\lim_{\lambda\to\lambda_c} b = 0$. The $\tilde\alpha$ and other
activated scaling exponents are described by the strong disorder universality 
fixed point in the renormalization sense. Extensive numerical simulations in 
\cite{Juhasz2011fk} provided estimates for them in case of GSW-s.

\section{Griffiths phases in Generalized Small World network models}
\label{sec:GSW}

\begin{figure}
 \centering
 \subfloat[]{
 \epsfxsize=55mm
 \epsffile{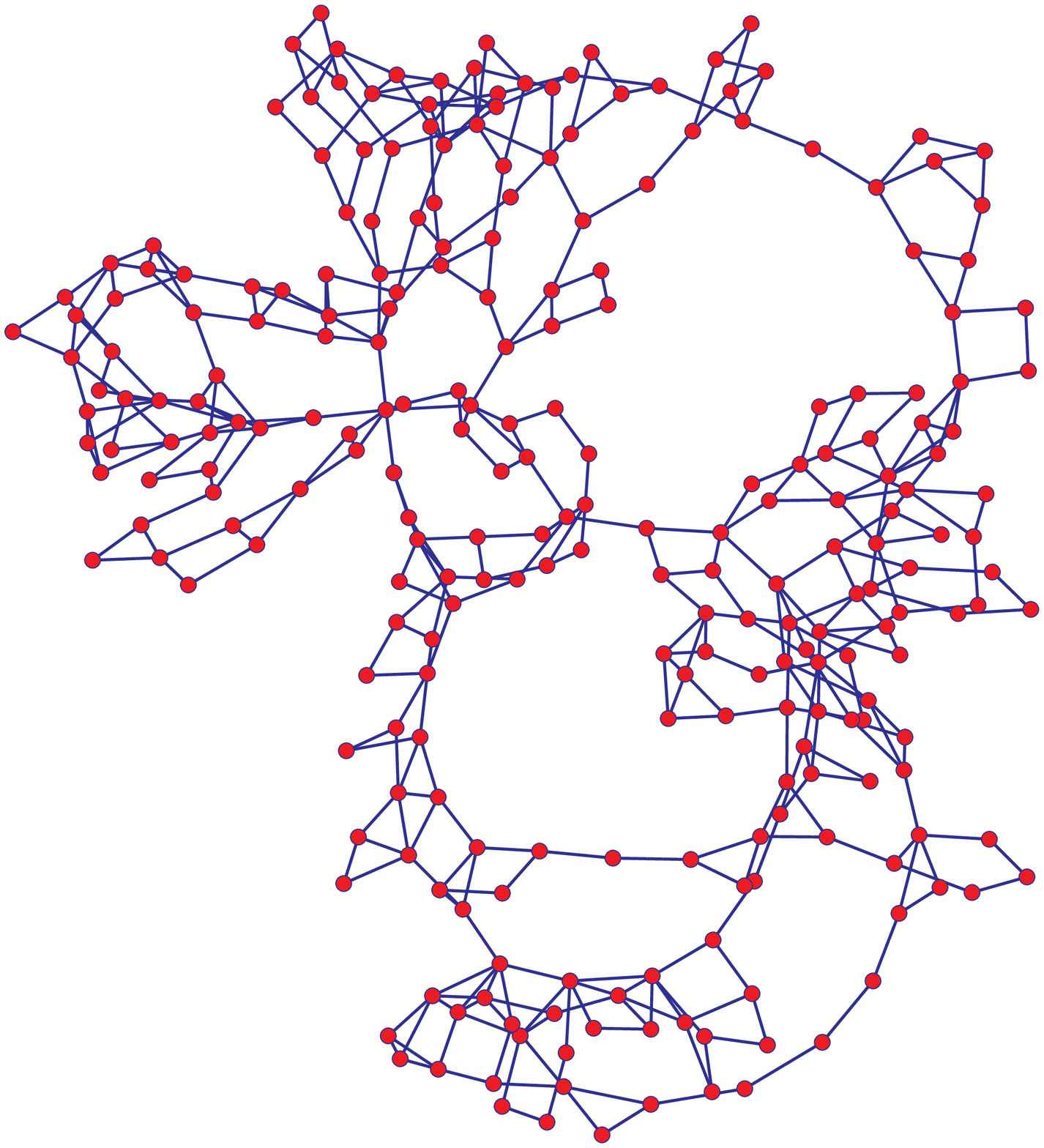}
 \label{fig:bb2-1}
 }
 \subfloat[]{
 \epsfxsize=60mm
 \epsffile{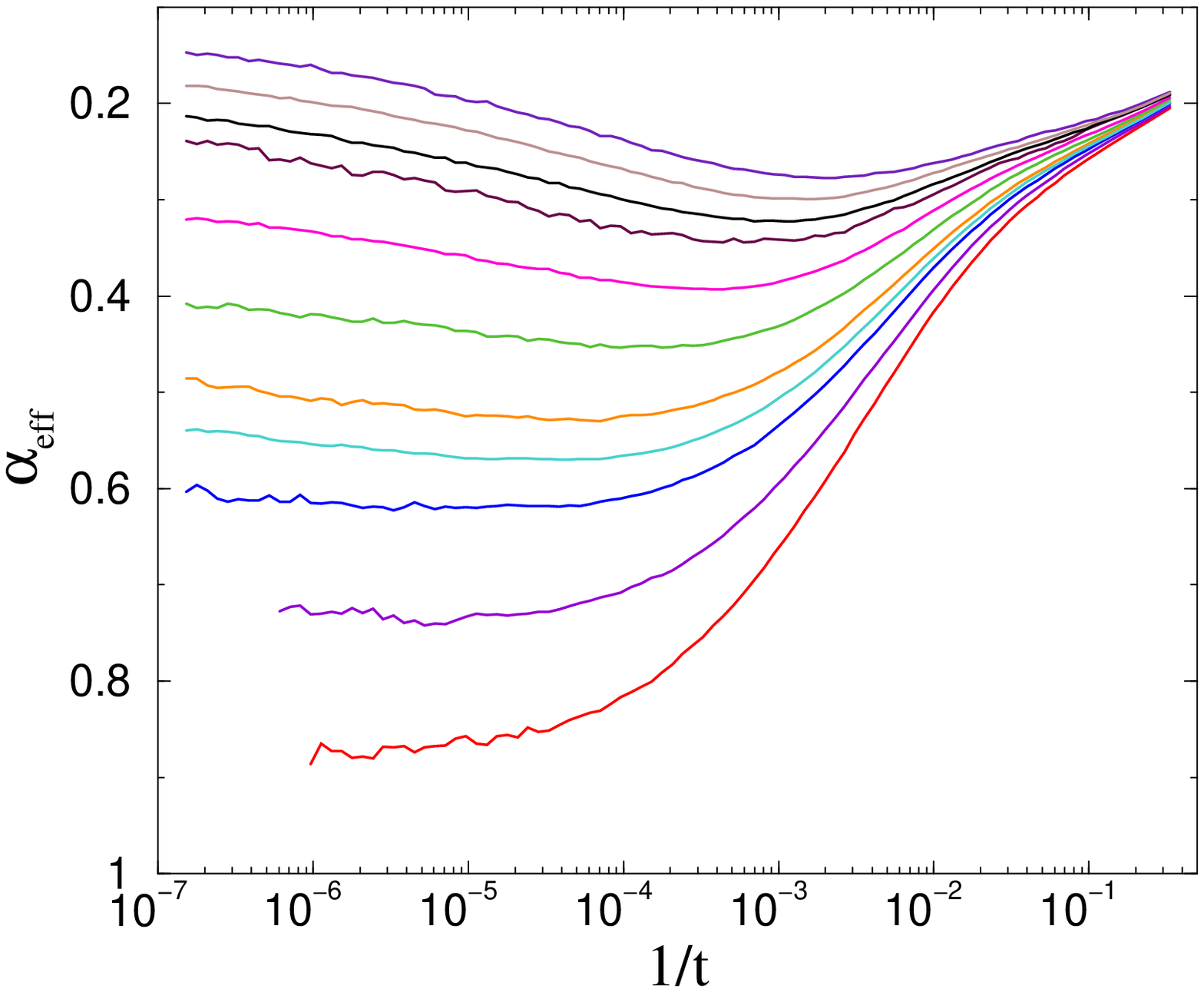}
 \label{fig:o2gi}
 }
 \caption{\subref{fig:bb2-1}~Topological view of a marginal GSW network 
 \cite{bb} with parameters $\beta=1$ and $N=256$ nodes.
 \subref{fig:o2gi}~Local slopes of the density decay in such networks with 
 $\beta=0.2$ and $N=10^5$ nodes.
 }
\end{figure}

GSW-s in one dimension have the intriguing feature that in the marginal case ($s=2$) 
intrinsic properties exhibit power-law behavior and the corresponding exponents 
vary continuously with the prefactor $\beta$. First it was shown that the
topological dimension of such networks (see Fig.~\ref{fig:bb2-1}) depends 
on $\beta$ \cite{bb,coppersmith,netrwre}.
It has been claimed in a recent letter \cite{GPCP} and in \cite{odor172,Juhasz2011fk} 
that if $d(\beta)$ is finite, GP-s and similar rare-region effects can also appear.
Density decay simulations, started from fully active state ($\rho(0)=1$),
have been presented in \cite{Juhasz2011fk}. The effective decay exponents,
defined as the local slope of $\rho(t)$:
\begin{equation}  \label{aeff}
  \alpha_{\rm eff}(t) = - \frac {\ln[\rho(t)/\rho(t')]} 
  {\ln(t/t^{\prime})} \ ,
\end{equation}
saturate to $\lambda$ dependent constant values in the long time limit 
(Fig.~\ref{fig:o2gi}), although logarithmic corrections arise.

Interestingly such GP-s have also been found in $3$-regular random networks,
constructed as shown \cite{Juhasz}. 
For larger values of $s$ the long-ranged links are irrelevant and $d(\beta)=1$, 
while for $s<2$ the topological dimension diverges and mean-field behavior emerges.

\section{Griffiths phases in Erd\H os R\'enyi network models}
\label{sec:ER}
  
The CP on Erd\H os R\'enyi (ER) graphs \cite{ER} with a quenched disordered 
infection rates (QCP) has been studied in \cite{GPCP,odor172,Juhasz2011fk}. 
A fraction $q$ of the nodes (type-II) propagate infection with a reduced value 
$\lambda r$, with $0 \le r < 1$, while the remaining fraction $1-q$ (type-I nodes) 
take their "clean" value $\lambda$.  
Pair mean-field approximations lead to the following critical threshold:
\begin{equation}
\lambda_{c}(q) = \frac{\langle k \rangle}{\langle k \rangle-1} ~
\frac{1}{1-q}.
\label{Crit}
\end{equation}
Type-I nodes experience a percolation transition, where the type~I-to-type~I
average degree is $1$, i.e. at $q_{perc}= 1-\langle k \rangle^{-1}$.  
For $q >q_{perc}$ activity cannot be sustained: type-I clusters are finite and
type-II ones do not propagate activity. Optimal fluctuation theory and simulations
show that in this case the absorbing phase of QCP splits and a GP with power-law
dynamics appears for $\lambda > \lambda_c^0$. 

\begin{figure}
 \centering
 \subfloat[]
 {
 \epsfxsize=50mm
 \epsffile{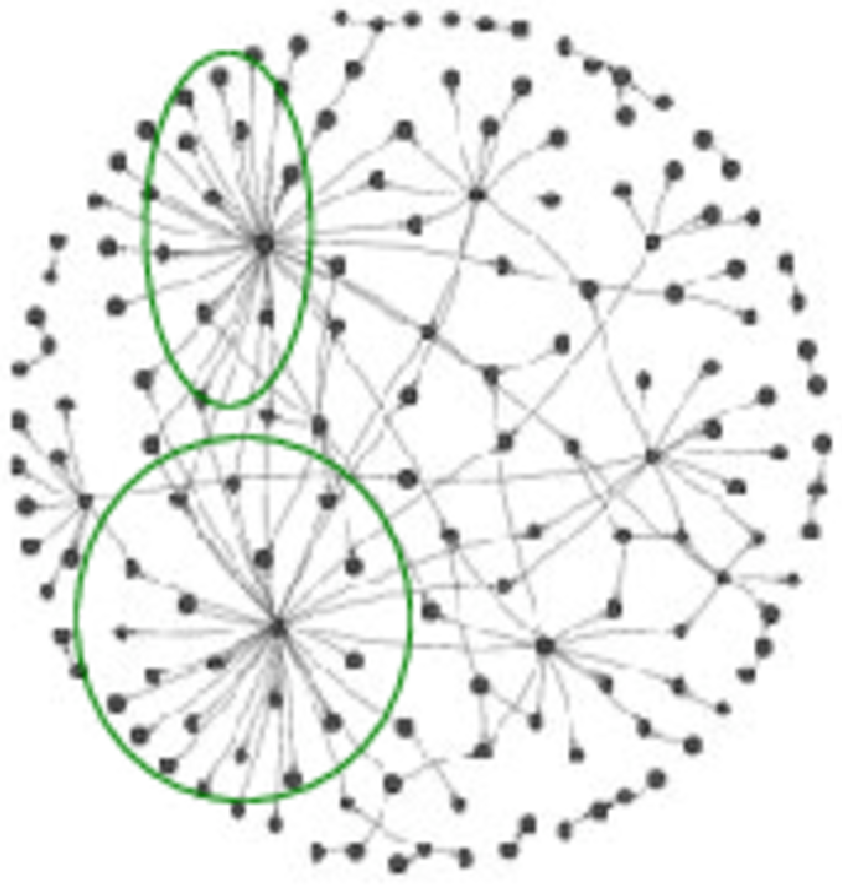}
 \label{fig:er}
 }
 \subfloat[]
 {
 \epsfxsize=60mm
 \epsffile{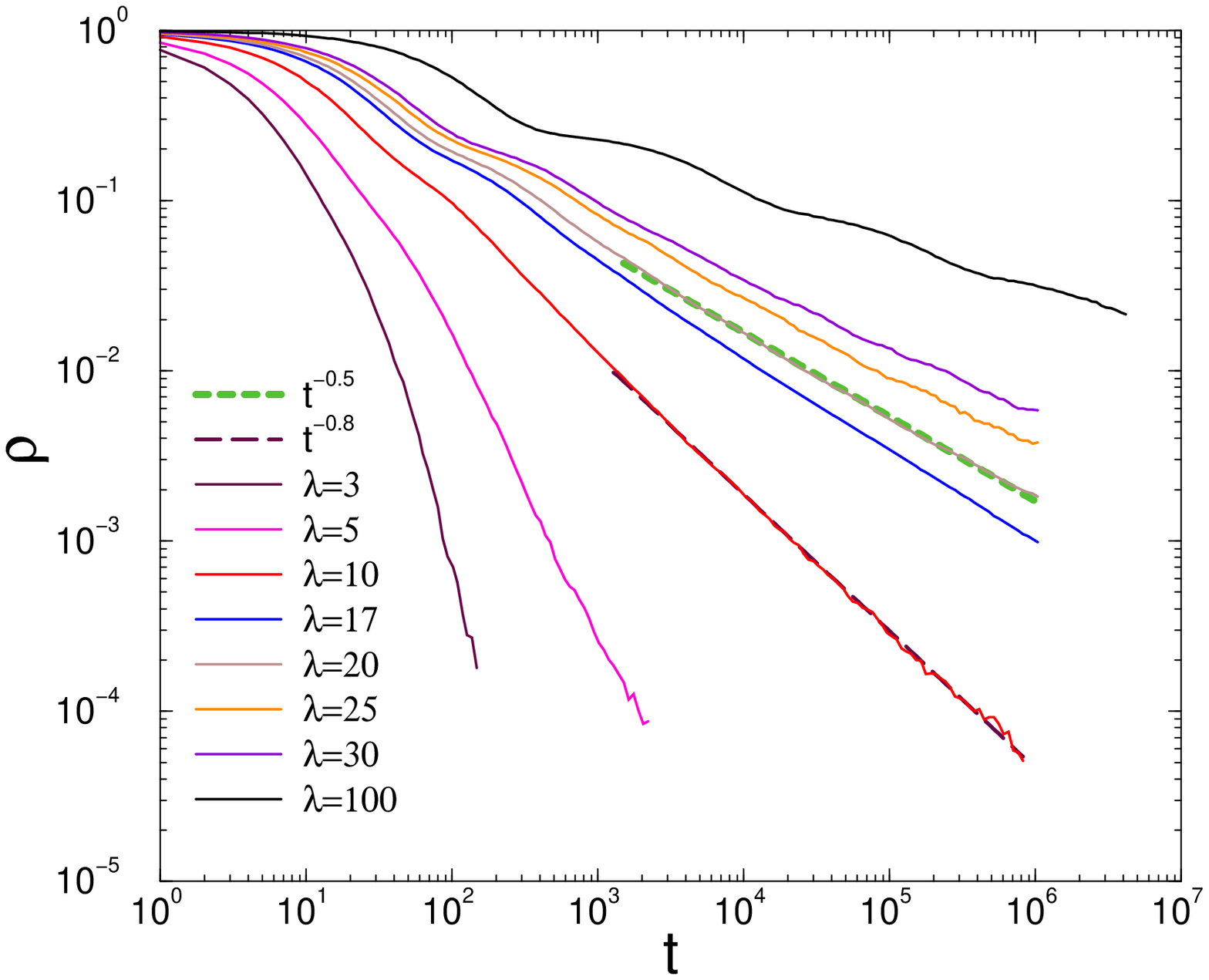}
 \label{fig:er03}
 }
 \caption{\subref{fig:er}~Picture of an ER network exhibiting RR-s.
          \subref{fig:er03}~Density decay of the CP in ER below the 
          percolation threshold. Dashed lines: power-law fits.}
\end{figure}

In fact in fragmented ER networks, where $d$ is zero the intrinsic disorder 
of $\lambda$-s is not necessary, but the topological heterogeneity itself 
is enough for the occurrence of GP-s as shown in Fig.~\ref{fig:er03}. 
Density decay simulations from fully active state of the CP on ER graphs
with $\langle k\rangle = 0.15$ up to sizes $N=10^6$ confirm this. 
For high values of $\lambda$ one can see plateaus on the $\rho(t)$ 
curves, similarly as reported in \cite{LSN12}.
These are the consequence of the metastable local active domains.

\section{Slow dynamics in Barab\'asi-Albert network models}
\label{sec:BA}

After showing the rare-region effects and the possibility of GP-s on 
GSW and ER networks let us turn towards the results on the most common type
of scale-free (SF) networks. CP have been simulated and analyzed by 
mean-field methods on BA networks, in particular for loop-less 
and weighted cases as described in \cite{BAGP}.  
BA construction is a simple and fast way to generate SF networks, in contrast with
other standard models, e.g \cite{ucmmodel}.
The BA growth starts with a fully connected graph of small ($N_0 = 10$) nodes.
Following that at each generation step $s$, a new vertex with $m$ edges 
is added to the network and connected to an existing vertex $s'$ of degree $k_{s'}$ 
with probability $\Pi_{s  \rightarrow s'} = k_{s'} /\sum_{s''<s} k_{s''}$.
This process is iterated until reaching the desired network size $N$. 
The resulting network has a SF degree distribution $P(k) \simeq k^{-3}$ and
for $m=1$ we obtain a BA tree (BAT) topology, while for the looped case
$m=3$ is used.
Binary (non-weighted) BA networks can be transformed into weighted
ones by assigning to every edge connecting vertexes $i$ and $j$ a
symmetric weight $\omega_{ij}$. In \cite{BAGP} two different
network topology dependent weight assignment strategy was introduced
in order to slow down and localize epidemics.

(i) \emph{Weighted BA tree I (WBAT-I):} Multiplicative weights, 
suppressing the infection capability of highly connected nodes
\begin{equation}
  \label{eq:2}
  \omega_{ij} = \omega_0 (k_i k_j)^{-\nu},
\end{equation}
where $\omega_0$ is an arbitrary scale and $\nu$ is a characteristic
exponent with $\nu\geq0$. This can model internal limitations of hubs,
like the sub-linear Heap's law \cite{Heap}.

(ii) \emph{Weighted BA tree II (WBAT-II):} Disassortatative weighting
scheme according to the age of nodes in the network construction
\begin{equation}
  \label{eq:3}
  \omega_{ij} =\frac{|i-j|^x}{N},
\end{equation}
where the node numbers $i$ and $j$ correspond to the time step
when they were connected the network. Since the degree of nodes 
decreases as $k_i\propto (N/i)^{1/2}$ during this process, 
this selection with $x>0$ favors connection between unlike 
nodes and suppresses interactions between similar ones.

\begin{figure}
 \centering
 \subfloat[]{
 \epsfxsize=60mm
 \epsffile{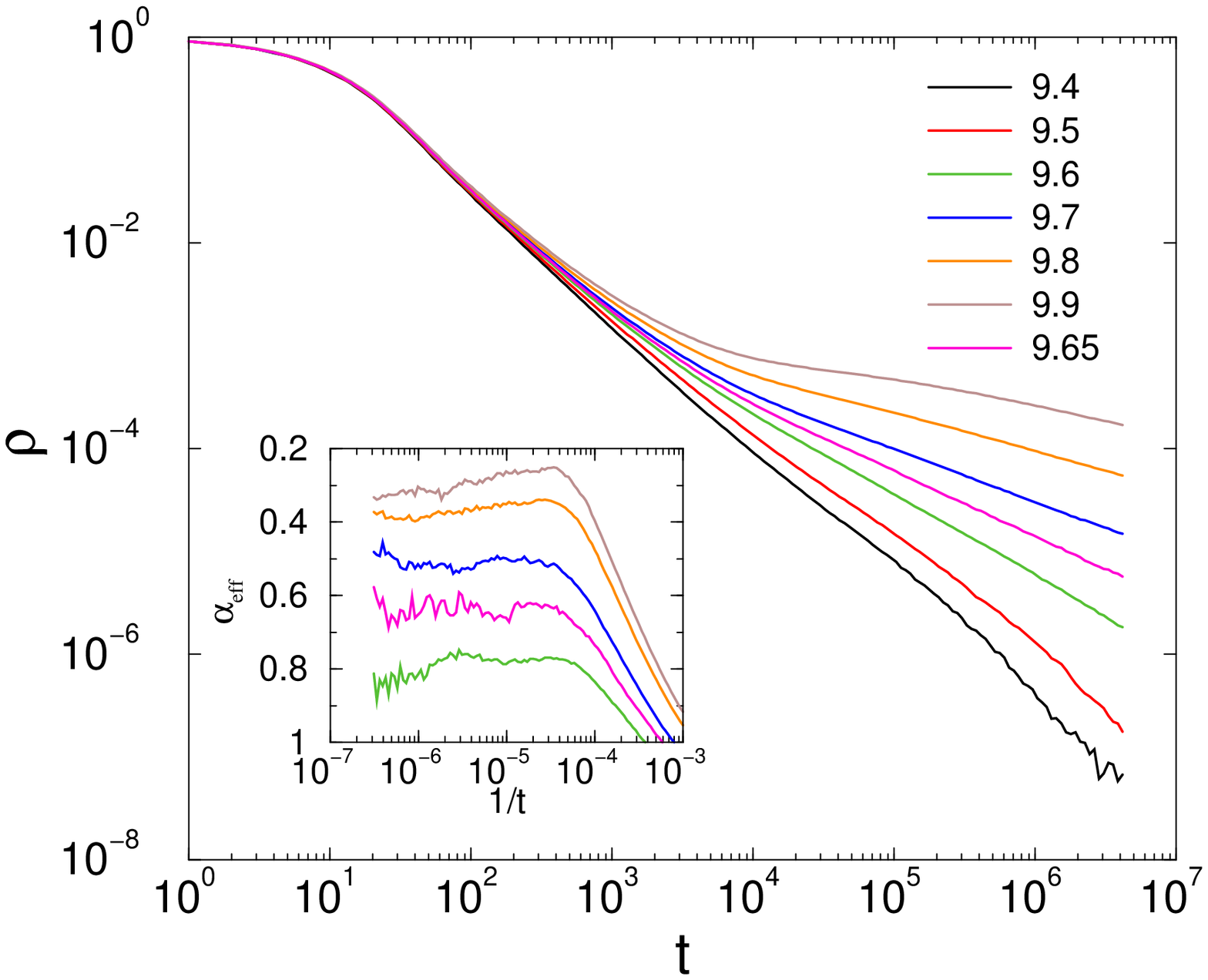}
 \label{gwbat2-3}
 }
 \subfloat[]{
 \epsfxsize=60mm
 \epsffile{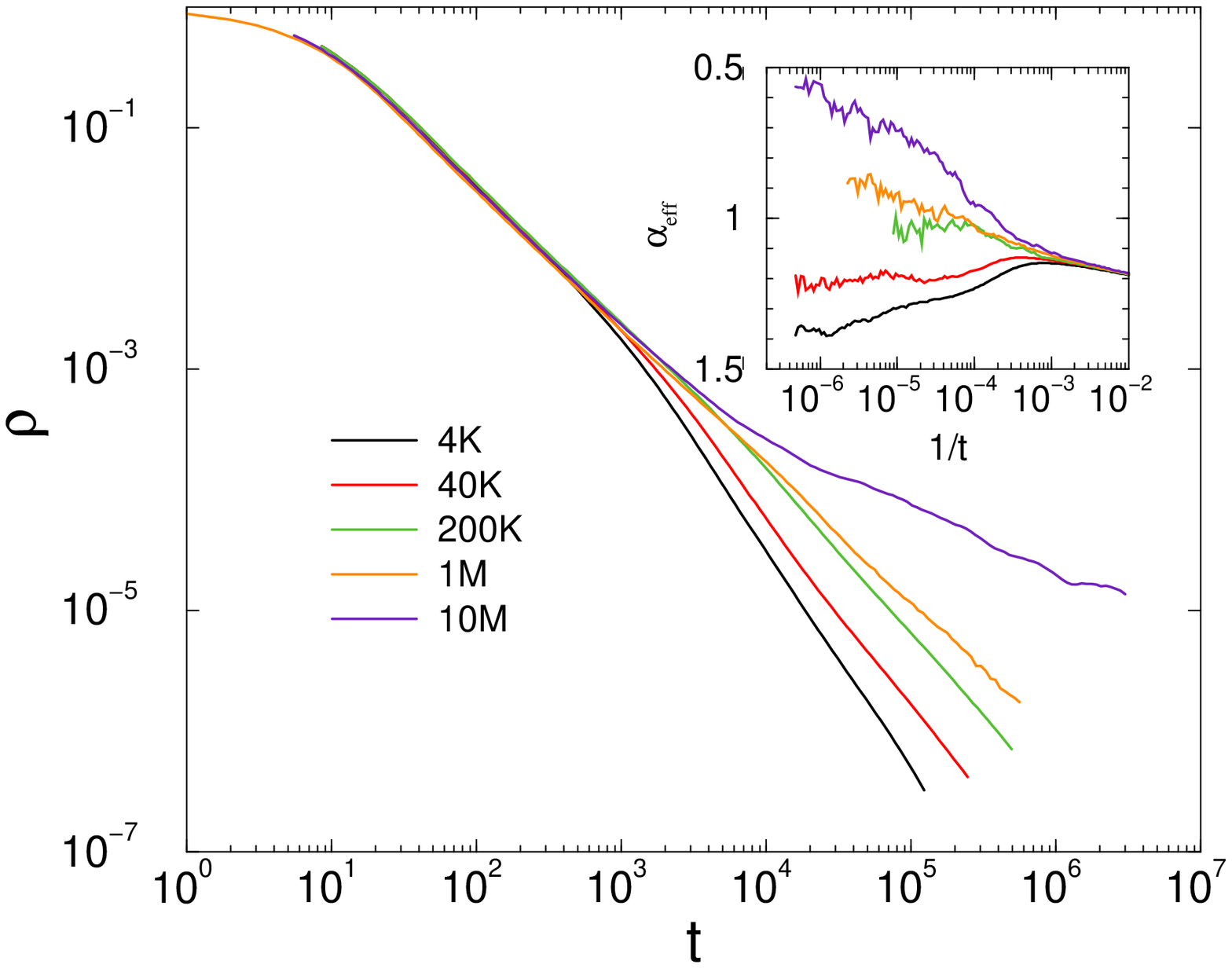}
 \label{gwbacp68}
 }
 \caption{\subref{gwbat2-3}~Density decay as a function of time in CP defined
  on WBAT-II trees, with exponent $x=3$. Network size: $N=4\times 10^5$. 
  Different curves correspond to: $\lambda=$ 9.4, 9.5, 9.6, 9.65, 9.7, 9.8, 
  9.9 (from bottom to top). Inset: the corresponding local slopes 
  \subref{gwbacp68}~Density decay as a function of time in CP on a weighted 
  WBAT-II trees with exponent $x=2$ at $\lambda=6.8$. 
  Different curves correspond to sizes $N=4\times10^3, 4\times10^4, 
  2\times10^5, 10^6, 10^7$ (from bottom to top).  
  Inset: the corresponding local slopes.
 }
\end{figure}

The presence of these weights affects the dynamics of the CP.
Thus, the rate at which a healthy vertex $i$ becomes ill on contact
with an infected (active) vertex $j$ is proportional to $\lambda \omega_{ij}$,
therefore the epidemic can in principle become trapped in isolated
connected subsets.
Density decay simulations \cite{BAGP}, started from fully active state 
show GP like regions both in the WBAT-I and WBAT-II cases. 
For WBAT-II the scaling appears for $t>10^4$ MCs in the region $\lambda>9.5$, 
even in case of networks with $x=3$ as shown on Fig.~\ref{gwbat2-3}. 
The effective decay exponents, using $t/t^{\prime}=8$, saturate to 
$\lambda$ dependent constant values in the long time limit 
(see inset in Fig.~\ref{gwbat2-3}). However, by increasing $N$ at a 
given $\lambda$ the decay curves saturate asymptotically, 
suggesting a smeared phase transition as shown on  Fig.~\ref{gwbacp68}.

\section{Discussion and Conclusions}

I have overviewed the effects of quenched heterogeneity on the dynamics of the
Contact Process in different network models. In finite topological dimensional
cases slow dynamics (power-law, logarithmic, stretched exponential ... etc) and
Griffiths Phases can be observed, which is very important for understanding
spreading-like phenomena of real world networks. In these cases topological
disorder of the network can result is GPs. Similarly, in models where 
the roles of space and time are exchanged “temporal Griffiths phases” have
been found in various systems near their phase transition points 
\cite{TGP1,TGP2}. 

On the other hand on graphs of infinite topological dimensions
like in weighted SF trees, the network heterogeneity can again cause 
slow dynamics of the CP. However, in the thermodynamic limit
the power-laws saturate and the locally active, high dimensional subspaces
cause smeared phase transitions \cite{BAGP}.
Very recently slow dynamics has also been reported in case of the 
Susceptible-Infected-Susceptible model defined on the (SF) flower model 
\cite{LSN12}, weighted BA trees \cite{wbacikk} and on ER graphs with 
exponential weights \cite{b13}. 
Spectral analysis of the quenched mean-field theory \cite{GDOM12, wbacikk} 
and simulations confirm epidemic localization effects in such models.
Further studies of different models and networks are currently under way.
 
\section*{Acknowledgments}

I thank R. Juh\'asz and I. Kov\'acs for their comments,
Romualdo Pastor-Satorras for discussions and hospitality
during my visits in Barcelona and acknowledge support from 
the Hungarian research fund OTKA (Grant No. T77629), 
HPC-EUROPA2 (pr. 228398) and the European Social Fund through 
project FuturICT.hu (grant no.:TAMOP-4.2.2.C-11/1/KONV-2012-0013).

\end{document}